\documentclass[conference]{IEEEtran}
\IEEEoverridecommandlockouts
\usepackage{cite}
\usepackage{amsmath,amssymb,amsfonts}
\usepackage{algorithmic}
\usepackage{graphicx}
\usepackage{textcomp}
\usepackage{xcolor}
\usepackage{orcidlink}
\usepackage{csquotes}
\def\BibTeX{{\rm B\kern-.05em{\sc i\kern-.025em b}\kern-.08em
    T\kern-.1667em\lower.7ex\hbox{E}\kern-.125emX}}
\begin{document}

\bibliographystyle{IEEEtran}

\title{Assessing the User Experience of Extended Reality Devices for (Dis)Assembly: A Classroom Study\\
}

\author{\IEEEauthorblockN{Brandon S. Byers}
\IEEEauthorblockA{\textit{Circular Engineering for Architecture} \\
\textit{ETH Zurich}\\
Zurich, Switzerland \\
\orcidlink{0000-0002-8622-8529}0000-0002-8622-8529}
\and
\IEEEauthorblockN{Eleftherios Triantafyllidis}
\IEEEauthorblockA{\textit{Circular Engineering for Architecture} \\
\textit{ETH Zurich}\\
Zurich, Switzerland \\
\orcidlink{0000-0001-7578-4290}0000-0001-7578-4290}
\and
\IEEEauthorblockN{Thibaut Menny}
\IEEEauthorblockA{\textit{Circular Engineering for Architecture} \\
\textit{ETH Zurich}\\
Zurich, Switzerland \\
\orcidlink{0000-0001-6113-6209}0000-0001-6113-6209}
\and
\IEEEauthorblockN{Martin Schulte}
\IEEEauthorblockA{\textit{Design Technologies} \\
\textit{Herzog \& de Meuron}\\
Basel, Switzerland \\
}
\and
\IEEEauthorblockN{Catherine De Wolf}
\IEEEauthorblockA{\textit{Circular Engineering for Architecture} \\
\textit{ETH Zurich}\\
Zurich, Switzerland \\
\orcidlink{0000-0003-2130-0590}0000-0003-2130-0590}
\and
\IEEEauthorblockN{   }
\IEEEauthorblockA{\textit{ } \\
}
}

\maketitle


\begin{abstract}
Despite the current rise and promising capabilities of Extended Reality (XR) technologies, the architecture, engineering, and construction industry lacks informed guidance when choosing between these technologies, especially for complex processes like assembly and disassembly tasks. This research compares the user experience across different XR devices for (dis)assembly utilizing the NASA Task Load Index and System Usability Scale metrics. Through a workshop and surveys with graduate civil engineering and architecture students, the study found that Augmented Reality scored highest in usability, followed closely by Mixed Reality. However, Mixed Reality showed the best task load index score, indicating low cognitive demand. The findings presented in this research may aid academics and practitioners in making informed decisions when selecting XR systems in practical, real-world assembly scenarios. Moreover, this study suggests opportunities and guidelines for more detailed XR system comparisons and exploration of XR's further role in circular construction practices.
\end{abstract}


\begin{figure*}[h]    
	\centering
  	\includegraphics[width=0.975\textwidth]{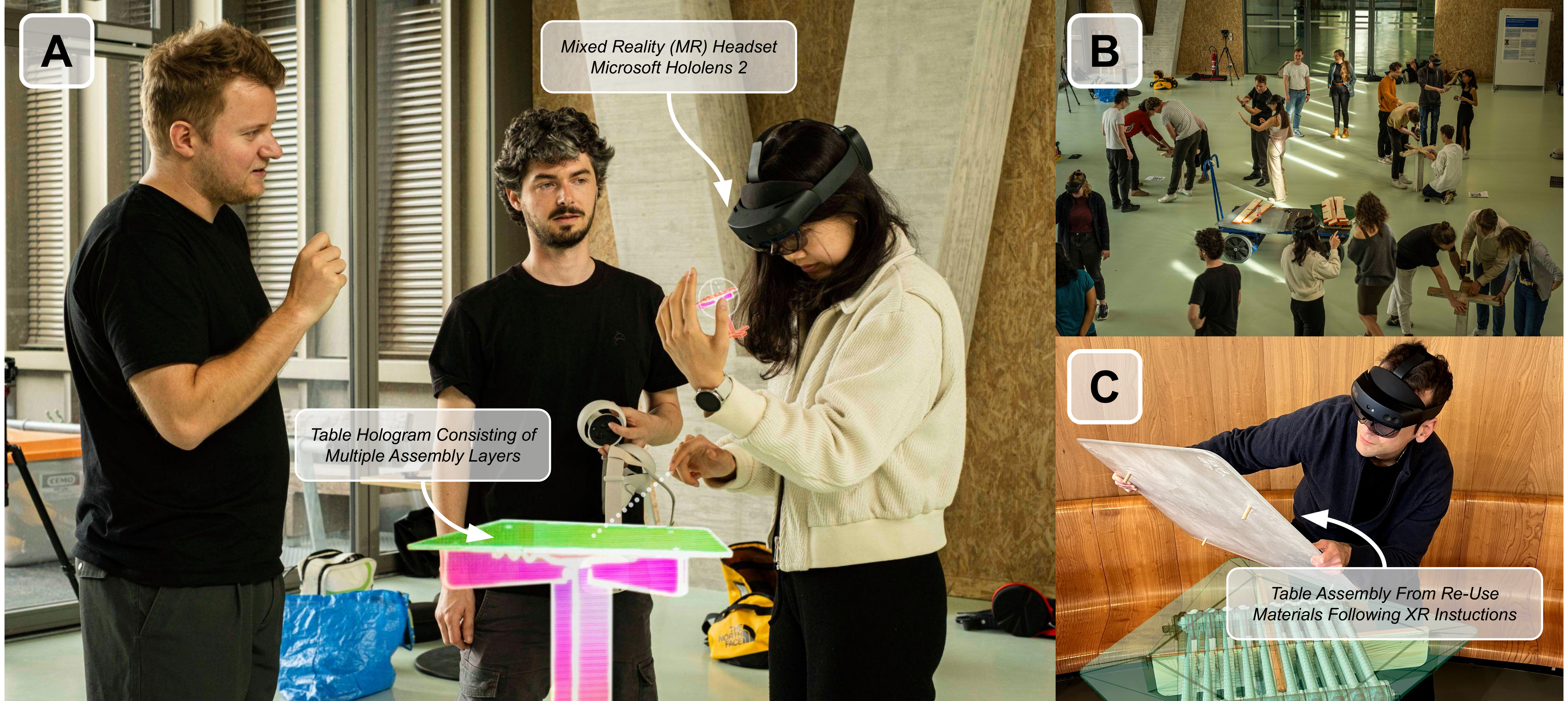}
        \vspace{3pt}
  	\caption{
  \textbf{Overview of using Extended Reality (XR) for the disassembly of tables made out of reused materials in a classroom study at ETH Zurich.}
  \textbf{(A)} User wearing the HoloLens with workshop facilitators.
  \textbf{(B)} Class overview of students using XR assisted technologies for step-by-step guided assembly.
  \textbf{(C)} Close-up of the real table with its digital model overlayed.
  }
  	\label{fig:workflow_diagram}
\end{figure*}

\section{Introduction}
The convergence of extended reality (XR) in architecture presents unprecedented opportunities to directly revolutionize and make a strong impact within the Architecture, Engineering, and Construction (AEC) industry \cite{de_wolf_d5_2024}. With the growth of XR and telecommunication technologies in recent years, the feeling of immersion and user experience has notably increased \cite{triantafyllidis_study_2020}, including for (dis)assembly processes \cite{mitterberger_augmented_2020}. Due to the potential of XR for providing highly realistic and immersive user experiences, digitally augmenting real-world spaces, coupled with data-enriched environments and real-time feedback loops, it has naturally received substantial focus in both research \cite{triantafyllidis_study_2020} and practical applications for AEC \cite{mitterberger_augmented_2020, mitterberger_extended_2023}. 

Nevertheless, from a practical perspective, especially within the AEC industry, there is a perceived risk of adopting new technologies for circular construction and digital buildings \cite{de_wolf_d5_2024}. Additionally, adoption of XR is slowed by inconvenient hardware, intimidated perceptions by new users, and learning curves for new technologies.

The goal of this classroom study was to provide participants with hands-on experience with state-of-the-art XR tools for sustainable building practices. In particular, this study demonstrates how XR can augment the intricate (dis)assembly of tables consisting of multiple reclaimed components with a real-time, step-by-step guided interface. The workshop aimed to highlight to students the potential of such novel tools in promoting circularity in industry.

\subsection{Background of XR in Practice}
XR applications in AEC practice can be classified into three primary categories: mobile-based augmented reality (AR), mixed reality (MR) headsets with pass-through, and fully immersive virtual reality (VR) systems, each offering distinct capabilities and limitations for professional use.

Mobile-based AR solutions have gained widespread adoption due to their accessibility through ubiquitous smartphone devices. These applications typically facilitate 360° virtual walkthroughs (e.g., \href{https://vrto.me/en/}{vrto.me}) and spatial 3D visualizations through platforms such as \href{https://sketchfab.com/}{Sketchfab}. While these tools effectively support client communication and public engagement in projects, they provide limited immersion compared to other XR modalities.

MR headsets have demonstrated particular efficacy through digital mock-ups, immersive design presentations across various scales (e.g., \href{https://iart.ch/en/next/aire}{iArt AIRE}), and even assembly guidance for timber construction projects (e.g., \href{https://timbar.ch/}{Timbar} and \href{http://fologram.com/}{Fologram}). The hands-free operation and physical space integration capabilities of MR systems present notable advantages. However, current MR technology faces several technical constraints, including restricted field of view, content optimization requirements, and gesture control interfaces that may challenge new users. Additionally, hologram drift and precision limitations can impact performance in assembly-focused applications.

While offering the highest degree of immersion, VR systems are primarily employed for internal design reviews and client presentations, either in person or through virtual collaboration platforms such as \href{https://www.arkio.is/}{arkio.is}. VR applications are divided into stand-alone applications, where the headset runs independently, and tethered applications, where the VR headset is connected to a workstation with a high-performance GPU. When connected to a high-performance workstation, VR headsets can visualize spaces in extremely high quality and are easy to use with software embedded into common CAD and BIM environments (e.g., \href{https://github.com/Unity-Technologies/unity-cloud-reference-project}{Unity Cloud Reference} and \href{https://irisvr.com/}{IrisVR}).
However, VR experiences can present challenges in use cases (for example, maintaining meeting continuity) due to user isolation, complex controls, and spatial orientation issues.

\subsection{Research on XR in AEC}
To use digital tools to facilitate a circular built environment and change construction processes, many technologies have been proposed. XR, as an umbrella term, is a promising suite of spatial computing technologies that have grown in interest and application to industrial environments, including construction. 

One of the earliest reviews of AR in AEC applications is from 2013 \cite{chi_research_2013}. The work emphasizes portable and mobile devices as being enabling technologies essential for providing information to workers in the field \cite{chi_research_2013}.
\cite{yu_applications_2024} conducted a systematic literature review on the challenges of adopting AR and VR technologies in construction. In their review of 24 publications, current applications involved education and training, auxiliary construction, and operation, maintenance, and safety. The challenges included latency, precision, comfort, safety, and quality. Future opportunities included using XR in collaboration with robots. The authors emphasize the need to focus on perceived utility for developing user-friendly applications \cite{yu_applications_2024}.

Using XR for assembly in construction has been explored before by \cite{pan_integrating_2023}. In their study, the researchers developed a framework combining smart construction objects, AR, and onsite assembly of housing modules \cite{pan_integrating_2023}. Though the work does take place in actual construction, it does not examine the user experience of the device. 
Similarly, \cite{pevec_-site_2024} used the HoloLens 2 and Trimble XR10 for overlaying a 3D Building Information Modeling (BIM) model to test compliance of the design model and as-built structure. After a survey of 25 participants, generally positive responses were recorded, stating that the device was very beneficial and welcome onsite \cite{pevec_-site_2024}.
\cite{li_integrating_2022} developed an approach to combine physical building models with a HoloLens XR device to enable renovation visualizations and improve public participation. Research on exploring XR applications for AEC is a flourishing field \cite{de_wolf_d5_2024}.

\subsection{Potential of XR for Circular Construction}
There are only a few examples of using XR devices for disassembly processes. Most of the relevant literature is derived from manufacturing, such as with dis/assembly path planning problems \cite{ghandi_review_2015} or AR for maintenance and repair strategies \cite{eswaran_augmented_2023}. One study examined the effectiveness of using AR for a pipe spool assembly task by comparing AR to a baseline (isometric drawing) \cite{kwiatek_impact_2019}. The participants were both professionals and students, the AR application was on an iPad, and the AR assembly instructions provided immense time savings, particularly with participants whose cognitive abilities measured as low \cite{kwiatek_impact_2019}.

An early study on the usefulness of VR for circular economy (CE) applications in construction was conducted by \cite{ogrady_circular_2021}. Their application used a VR model of a real building called Legacy Living Lab, and the virtual BIM allows users to see different parts and properties of the prefabricated building \cite{ogrady_circular_2021}. 
Similarly, \cite{sahebzamani_integrated_2023} developed a BIM and VR workflow to visualize different CE scenarios for stakeholders through OneClickLCA data, although there was no testing or validation.
\cite{parry_recycling_2021} used an MR interface for inventorying, cutting, and reusing wood scrap in a new bespoke design for stacking timber.

As part of the D5 Digital Circular Workflow, XR was proposed for disassembly and deployment actions, but specific XR devices were not compared \cite{de_wolf_d5_2024}. \cite{soman_extended_2024} examined an array of XR applications for a circular economy, providing research directions and some cases studies for narrow, slow, close, and regenerate strategies as well as for building life cycle stages. For end-of-life applications, examples are provided on using AR for maintenance interventions and selective disassembly planning for buildings, but there is only one documented disassembly application, which is for a gearbox \cite{soman_extended_2024}. Similarly, XR is proposed as a digital technology for driving circular economy by \cite{chi_driving_2023}, but there is no guidance or testing to support the claim.

\subsection*{Research Aim}
Ultimately, XR is often proposed as an approach to aid in the construction and deconstruction processes in a circular economy, nevertheless, there is very limited research on using XR for disassembly applications, nor is there literature comparing XR devices in the AEC industry.
It is not well known how potential users would rate the experience and confidence of using different XR systems. 

Therefore, the objective of this research is to better understand the user experience for XR devices in the application for building product disassembly. In this study, user experience is defined as a combination of the System Usability Score (SUS) and NASA Task Load Index (TLX). This objective is explored through the following research question: \enquote{\textit{\textbf{How does the user experience vary between AR, MR, and VR devices for a real-life hands-on (dis-)assembly task?}}} 

\section{Methods}

This exploratory research is conducted within a pragmatist research paradigm that focuses on practical applications and outcomes from the experiment. The cross-sectional experiment uses quantitative methods by issuing a survey to participants after completing the experimental work.

\vspace*{-6pt}
\begin{figure}[tb] 
	\centering
	\includegraphics[width=0.7\linewidth]{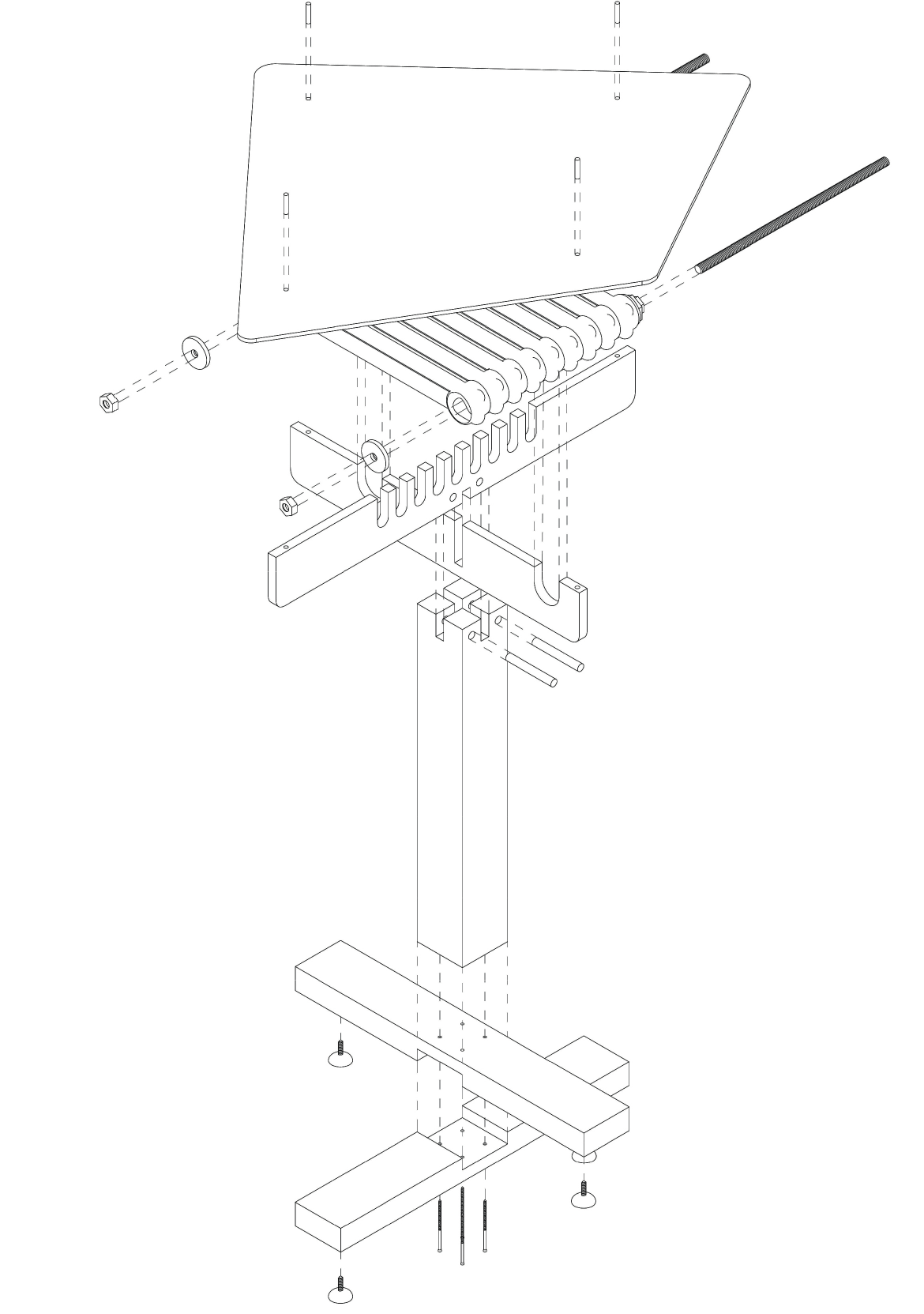}
	\vspace{5pt}
        \caption{AdapTable exploded view.}
	\label{fig:table}
\end{figure}

\subsection{The Experiment}
The experiment was a semi-structured classroom-based workshop held in the Spring of 2024. The workshop was conducted via graduate level course work in civil engineering and architecture at ETH Zurich in collaboration with the firm Herzog \& de Meuron. The participants were broken into two groups consisting of four sub-groups each to run the workshop in two sessions. Each sub-group was assigned to a station that had a table designed for disassembly, which was called AdapTable as seen in Figure \ref{fig:table}). 

The workshop was designed to require disassembly and re-assembly of the AdapTables using various XR devices. The semi-structured task is designed to enable curiosity, play, and individual experimentation with the devices. Before the workshop, participants were requested to download Rhino8 and the Fologram application for their smartphone or tablet. The initial 3D model of the table is provided with the disassembly layers for downloading into the AR environment in Fologram. For the MR study, Microsoft HoloLens devices are provided with the model and disassembly layers. Lastly, Meta Quest 2 devices are used for the VR category, with the model pre-loaded onto the device.

The task in the workshop is for each group to disassemble the table while one of the group members wears a HoloLens device and directs the other members. The members in the groups are permitted to swap participants using the MR device.
After the initial disassembly task was completed, the groups are requested to reassemble the table using the AR and VR devices. At this stage, the students are allowed to switch between any device. 
The AR and MR devices allow for a live reference model, while the VR device requires referencing the model in the device and needing to remove it for directing reality.
The semi-structured nature of the experiment limits the direct comparability of the devices and tasks but nonetheless facilitates open exploration of the XR devices (Figure \ref{fig:workflow_diagram}).



\begin{figure*}[h]    
	\centering
  	\includegraphics[width=0.825\textwidth]{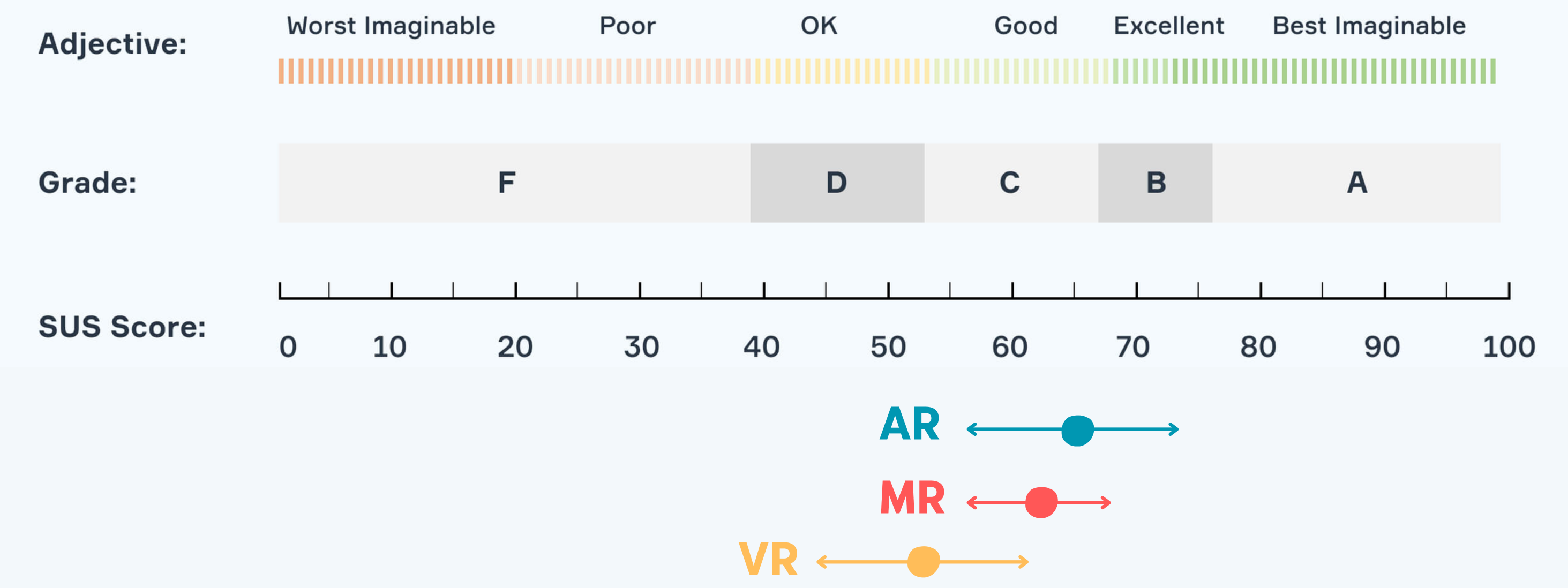}
        \vspace{3pt}
  	\caption{SUS Average Scores of the Devices with Confidence Interval (Initial image from \cite{sauro_5_2018} and adapted by authors).}
  	\label{fig:SUS_Scores_wCI}
\end{figure*}

Following the disassembly and assembly exercises the groups switched out and were asked to fill out an optional survey at the end. The survey was through Google Forms and consisted of a combination of the SUS \cite{brooke_sus_1996}, NASA's TLX, and questions on prior familiarity with the devices. The familiarity questions were asked on a five-point Likert scale, with five being very familiar. The TLX was asked on a ten-point Likert scale, with ten being very high [cognitive load]; and the SUS was asked on a ten-point Likert scale, with ten being strongly agree.
Given the open nature of the workshop and surveys, an unequal distribution of responses was collected. In total, 35 responses were received on the XR devices, consisting of three for AR, seven for VR, and 25 for MR.

\subsection{The Analysis}
The final stage of the methods is the analysis of the data. This study aims to explore and compare the user experience between XR devices for a (dis)assembly task. The user experience is assessed by both the SUS and TLX scores on the particular device tested. These both have precedence by being used before for similar experiments for user ratings. The averages, precision, and confidence levels are reported for each set of scores. Based on the collected sample size (\textit{n}), calculated standard deviation (\textit{s}), t-value (\textit{t}) derived from the confidence interval (90\%) and degrees of freedom (\textit{n-1}), the margin of error (\textit{d}) can be calculated as derived from the sample size determination equation (\ref{eq:01}).

\begin{equation}\label{eq:01} 
  n = \frac{t^2 s^2}{d^2}
\end{equation}

Given the unbalanced and underpowered sample size, this research does not attempt to state or reject hypotheses that one XR system has a better user experience than the other, nor that using an XR system is better than no system. Instead, the exploratory analysis of this work aims to collect user experience data on different devices to provide a directional indication for future user studies and practitioners interested in adopting XR devices.

\section{Results}
\begin{figure}[htbp!] 
	\centering
	\includegraphics[width=1.0\linewidth]{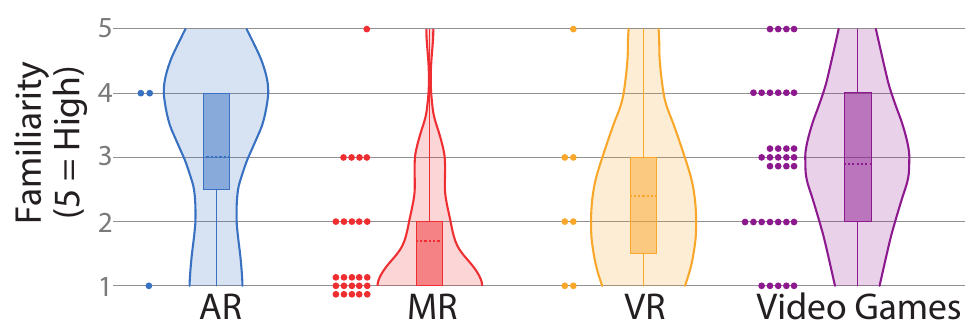}
        \vspace{3pt}
	\caption{Violin plot distribution of the participants familiarity with each XR system and video games (for reference).}
	\label{fig:fam}
\end{figure}

A total of N=35 responses were collected for all XR devices from the workshop. The first set of questions asked the participants to rate their experience with the device they are about to rate. Next, the survey asked about their familiarity with video games to reference against the XR devices (see Figure \ref{fig:fam}). The responses show the average familiarity for AR is 3.0/5.0, for MR is 1.7/5.0, for VR is 2.4/5.0, and for video games is 2.9/5.0.

\vspace*{-6pt}
\nointerlineskip 

\subsection{Result I: SUS Analysis}

\begin{figure}[!htbp] 
	\centering
	\includegraphics[width=1.0\linewidth]{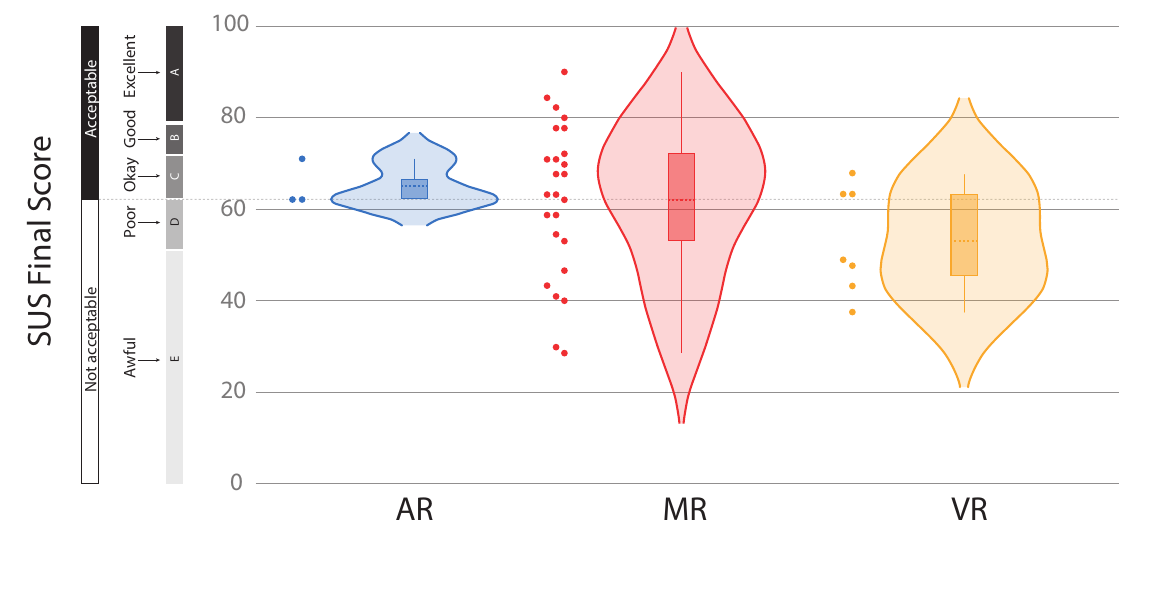}
	\caption{Violin plot of SUS distribution results per XR device.}
	\label{fig:SUS}
\end{figure}
\nointerlineskip 

The results for the SUS analysis are reported below and visualized in Figure \ref{fig:SUS_Scores_wCI} as well as in Figure \ref{fig:SUS}.
For the AR device (n=3), the average SUS is 65.2, which is marginally high or nearly good on the SUS scale. There is 90\% confidence that the true mean SUS for the AR device is within 56.5 to 73.8.

For the MR device (n=25), the average SUS is 62.3, which is marginally high and between ok to good on the SUS scale. There is 90\% confidence that the true mean SUS for the MR device is within 56.6 to 68.0.

For the VR device (n=7), the average SUS is 53.2, which is marginally low or only ok on the SUS scale. There is 90\% confidence that the true mean SUS for the VR device is within 44.7 to 61.7.

\subsection{Result II: NASA TLX Analysis}
\begin{figure}[!htbp] 
	\centering
	\includegraphics[width=0.9\linewidth]{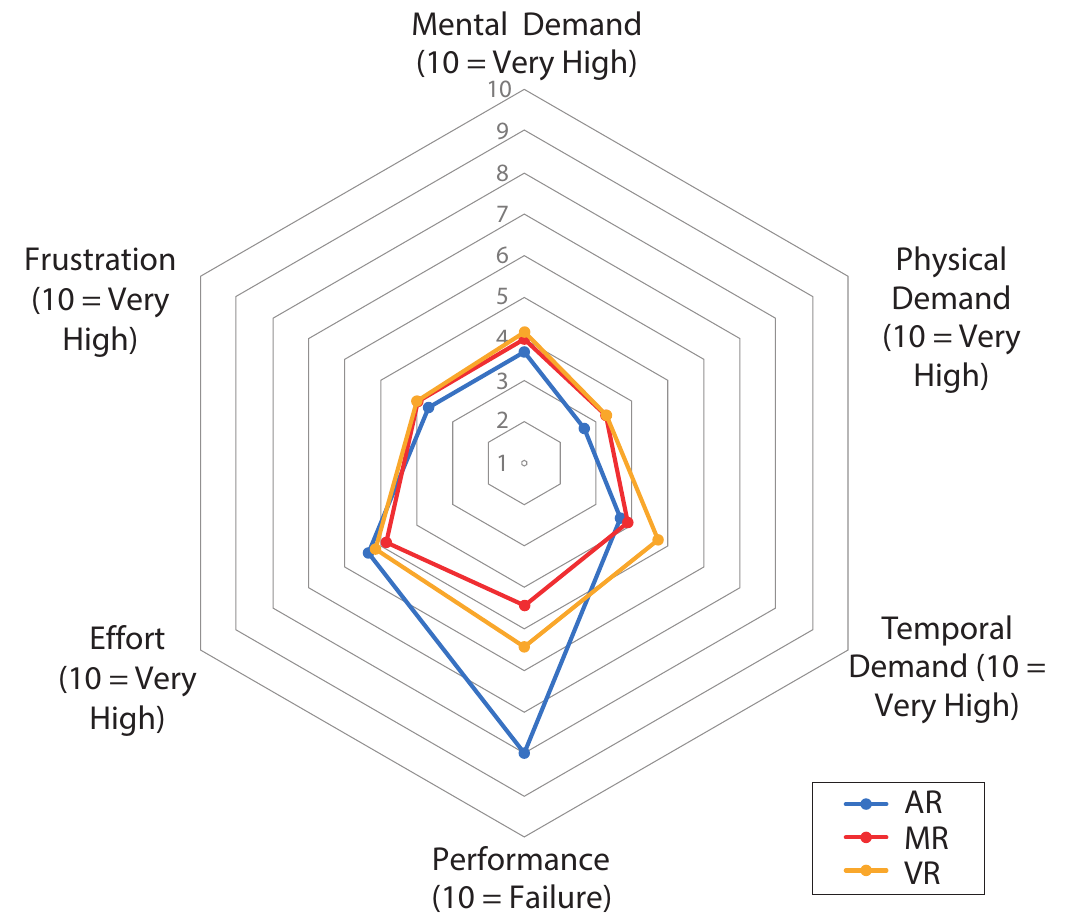}
	\caption{Radar plot for each TLX response per XR device.}
	\label{fig:TLX}
\end{figure}

The results for the TLX analysis are reported below and visualized in Figure \ref{fig:TLX}. 
The lower TLX score indicates a lower task load for the user.

For the AR device (n=3), the average TLX is 4.5 / 10, which is low on the TLX scale. There is a 90\% confidence that the true mean TLX for the AR device is within 3.3 to 5.7, due to large standard deviation and small sample size.

For the MR device (n=25), the average TLX is 4.1 / 10, which is moderately low on the TLX scale. There is a 90\% confidence that the true mean TLX for the MR device is within 3.7 to 4.5, due to the low standard deviation and large sample size.

For the VR device (n=7), the average TLX is 4.5 / 10, which is moderately low on the TLX scale and in accordance with other work from \cite{triantafyllidis_study_2020}. There is a 90\% confidence that the true mean TLX for the VR device is within 3.9 to 5.1.

\nointerlineskip 

\section{Discussion}
When examined on an absolute scale, the SUS scores do not rank very high, indicating that the devices are still not optimized for the task provided. 
Overall, the AR device exhibits the highest SUS, which is a promising result, yet limited by the small sample size and, therefore, a large margin of error. The MR device had the best TLX with the smallest margin of error, yet scored only moderately in the SUS. The VR device had the worst combined user experience, albeit with a similarly large margin of error. Naturally, this indicates the need for larger sampling for greater confidence. Nevertheless, due to the exploratory nature of the workshop, the results are directionally interesting. 

Several key challenges constrain XR adoption into practice. Current device form factors impact user comfort during extended use, while novel virtual environments require substantial training for non-intuitive controls, orientation, and interfaces. Unlike web design, XR lacks standardized interaction conventions that users can intuitively understand.
Technical constraints further complicate implementation. XR data preparation often adds time-consuming steps to project workflows and requires different formats than traditional planning processes. These challenges, combined with difficulties in quantifying return on investment, present significant adoption barriers.

Although this research was focused on a small-scale and controlled (dis)assembly exercise, the presented findings are useful. They are the first known contribution to the literature of XR in AEC that focuses on comparative user experience. These findings can inform decision-makers, both those designing research studies and practitioners who want to choose an XR device for implementation. These findings supplement existing research that calls for the use of XR for (dis)assembly applications without providing guidance on device selection \cite{soman_extended_2024,de_wolf_d5_2024}.

\subsection{Limitations}
Although this is an exploratory study using a semi-structured workshop, several limitations are inhibiting the applicability of the experiment.
Firstly, this is a lab-based experiment not representative of an actual environment in the field. Similarly, the population of workshop participants, though varied in background and professional interests, is not representative of the AEC population. Because of the open nature of the workshop, an unbalanced amount of responses were received per device limiting significant comparison. \textit{The results between devices should not be compared one-to-one}. The (dis)assembly task was relatively simple, especially compared to \cite{kwiatek_impact_2019}, and could be developed further into more complicated tasks. This study does not cover efficiency nor baseline testing against a control without an XR device. The form factor and often non-intuitive software are also potential limitations for adoption. Additionally, there is potential for spatial and modeling errors in the devices. Nevertheless, the limitations provide grounds for future work, which leaves a lot for additional exploration. 

\subsection{Call for Future Work}
This work is a preliminary study on the comparison between XR devices for a circular construction application. On account of the specific scope, research design, and experimental limitations, there are many ways to extend this research.
If XR is to be used to facilitate circular construction, then studies should develop a baseline on the efficiency and user experience of an XR device relative to a traditional approach (e.g., with paper drawings) like \cite{kwiatek_impact_2019}. Similarly, streamlined data processing pipelines between CAD/BIM and XR platforms would help minimize workflow disruptions.
Exploring this in a field context, similar to \cite{pan_integrating_2023}, can provide more nuanced insights into the potential advantages of using XR.

Nonetheless, it is important to not only account for technology-related tracking inaccuracies that can negatively affect user performance \cite{triantafyllidis_metrics_2022}, but also the accumulation of spatial errors over time when considering human-computer interaction is of importance \cite{triantafyllidis_advancements_2024}. The implementation of automatic spatial positioning systems, such as object snapping to physical elements, could enhance precision and usability in mixed reality applications. 

From a biological standpoint, users tend to utilize a wide range of stimuli when perceiving their surrounding environment \cite{triantafyllidis_advancements_2024,turk_multimodal_2014,ghazanfar_is_2006}. While XR offers an immersive audio-visual experience, coupling it with other sensory stimuli, most prominently tactile sensory feedback, has shown to notably enhance user immersion and performance in complex 3D tasks \cite{triantafyllidis_study_2020,triantafyllidis_advancements_2024,sigrist_augmented_2013}. 

Finally, while XR is a promising tool, designing intuitive user interfaces that take into account the intricate human biomechanical motions associated with ergonomics is considered crucial according to scholars \cite{triantafyllidis_advancements_2024,triantafyllidis_metrics_2022,triantafyllidis_challenges_2021}. The development of more intuitive control systems, potentially leveraging artificial intelligence, could significantly reduce the learning curve for new users.

\section{Conclusions}
This research compared user experiences across different XR devices in a (dis)assembly workshop, using the SUS and NASA TLX as metrics. While there is existing research on XR applications for AEC, no researchers have compared device user experience, particularly within circular construction contexts.

The study revealed several key findings:

\begin{enumerate}
    \item AR devices scored highest in usability but showed a high task load and had the lowest precision due to a small sample size
    \item MR devices achieved comparable usability scores with lower task load and higher precision
    \item VR devices matched AR's cognitive demands but scored significantly lower in usability
\end{enumerate}
When considering both usability and cognitive load, MR and AR devices emerged as better options. However, the AR results demonstrate lower statistical reliability due to their larger margin of error compared to the MR findings.

These insights offer immediate value to practitioners implementing XR technologies and establish benchmarks for future research on XR device performance and comparison. As the AEC industry increasingly adopts digital technologies, these findings could contribute to more effective and sustainable construction practices.

\section*{Acknowledgment}
Eleftherios Triantafyllidis is supported by the Albert Lück Stiftung via the ETH Zürich Foundation for the project \enquote{\textit{Extended Reality for Inspection, Assembly, Operations for Net-Zero Carbon Infrastructure}}.
The authors thank the course students for participating, to Herzog \& de Meuron for supplying us with the necessary devices, for Elias Knecht's overall help with the course and photography, and, finally, to Anna Buser
for supplying us with the photography content presented in this paper.

\bibliography{XRreference}

\vspace{12pt}

\end{document}